\documentclass{nature}

\usepackage{CJK} 
\usepackage[utf8]{inputenc}
\usepackage[T1]{fontenc}
\usepackage{amsmath}
\usepackage{color}
\usepackage{graphicx}
\let\realincludegraphics\includegraphics
\AtBeginDocument{\let\includegraphics\realincludegraphics}
\makeatletter
\renewenvironment{figure}{\@float{figure}}{\end@float}
\renewenvironment{figure*}{\@dblfloat{figure}}{\end@dblfloat}
\renewenvironment{table}{\@float{table}}{\end@float}
\makeatother
\usepackage{ulem}
\usepackage{longtable}
\usepackage{booktabs}
\usepackage{lineno}
\linenumbers

\nolinenumbers
\spacing{1.2}

\title{
Model-free Reconstruction of Molecular Energy Levels by Broadband Kilohertz-accurate Cavity-enhanced Spectroscopy
}

\author{S. Vasilchenko$^{1,\dagger}$, A.-W. Liu$^{1,3,\dagger,*}$, C.-X. Zuo$^{2}$, Y.-Q. Cheng$^{3}$, Z.-T. Zhang$^{2}$, W.-T. Wang$^{2}$, Y. R. Sun$^{4}$, Y. Tan$^{2}$, and S.-M. Hu$^{1,2,*}$}

\begin{document}
\begin{CJK*}{UTF8}{gbsn} 

\flushbottom
\maketitle

\begin{affiliations}
\item Hefei National Laboratory, University of Science and Technology of China, Hefei 230088, China
\item Hefei National Research Center for Physical Sciences at the Microscale, University of Science and Technology of China, Hefei 230026, China
\item State Key Laboratory of Chemical Reaction Dynamics, Department of Chemical Physics, University of Science and Technology of China, Hefei 230026, China
\item Institute of Advanced Light Source Facilities, Shenzhen 518107, China
\end{affiliations}

\noindent$^{\dagger}$These authors contributed equally to this work. \\ %\hspace{1em} 
\noindent $^{*}$Correspondence to: awliu@ustc.edu.cn; smhu@ustc.edu.cn.

\vspace{0.2in}

\begin{abstract}
Assigning the lines of high-resolution molecular spectra to quantum states requires a Hamiltonian model and substantial expert intervention, so the spectra of larger molecules accumulate vast numbers of unassigned lines. Here we show that molecular energy levels can instead be reconstructed directly from the raw, unassigned transition frequencies, using graph theory alone with no model and no prior assignment. Our inverse graph construction exploits recurring frequency differences and four-cycle closures to assemble an energy-level network at kilohertz precision. The dense, broadband spectra this requires are produced by a cavity-enhanced spectrometer (SCALS) that scans continuously across tens of terahertz at kilohertz accuracy, combining broadband coverage, high sensitivity, and high precision in a single automated instrument. Applied to the water absorption spectrum in the range of 1537--1605~nm, 686 Lamb dips of water were obtained without assignments, and the method reconstructs 158 energy levels that are mostly two orders of magnitude more precise than the corresponding literature values. By removing the assignment barrier, this approach opens a route to exploratory precision spectroscopy of polyatomic molecules without a priori knowledge of transition frequencies.
\end{abstract}

\thispagestyle{empty}

\newpage
% ============================================================
Precision spectroscopy of narrow optical transitions underpins modern metrology~\cite{Quinn2003Met, Dimarcq2024Met-second, Leung2022PRX-2P-Sr2, Yamaguchi2024Nature-Th, Zhang2024Nature-Th}, tests of fundamental physics and of new physics beyond the Standard Model~\cite{Grinin2020Science-2P-H, Derevianko2014NP-DarkMat, Biesheuvel2016NC-HD, Ubachs2016RMP, Tao2018PRL, Safronova2018RMP}, and provides indispensable reference data for spectroscopic databases~\cite{GORDON2026HITRAN2024}, frequency standards~\cite{BIPM2024_C2H2, Madej2006JOSAB}, communications~\cite{Riehle2018Met}, astrophysics, and atmospheric science~\cite{Vasilchenko2023JQSRT}. Yet the reach of precision spectroscopy has long been limited less by the ability to measure than by the ability to analyse: assigning observed lines to quantum states is a model-driven, iterative procedure, in which an effective Hamiltonian is fitted and refined against the data with substantial expert intervention. Even for methane (CH$_4$), a five-atom molecule that is by no means the most complex, the extreme density of rovibrational lines produced by strong rotation-vibration interaction has made its spectroscopic analysis a formidable, decades-long challenge: despite sustained effort, a significant fraction of the measured lines in even its most important windows remain unassigned~\cite{Nikitin2017JQSRT-CH4, Rey2018Icarus-CH4-Titan}. This assignment barrier, rather than the measurement itself, is often what prevents precision spectroscopy from being extended to larger and more complex molecular systems.

Graph-theoretic frameworks, such as MARVEL~\cite{Furtenbacher_MARVEL_JMS} and the broader spectroscopic-network approach~\cite{Csaszar2011JMS-SN, Csaszar2016, Tobias2020, Diouf2021JPCRD-H2O}, invert measured transitions into empirical energy levels with well-defined uncertainties and now underpin modern line-by-line databases. These methods are, however, assignment-driven: they operate on transitions that already carry quantum-number labels, and so presuppose precisely the step that remains the bottleneck for larger molecules.

\begin{figure}
  \centering
  \includegraphics[width=0.5\textwidth]{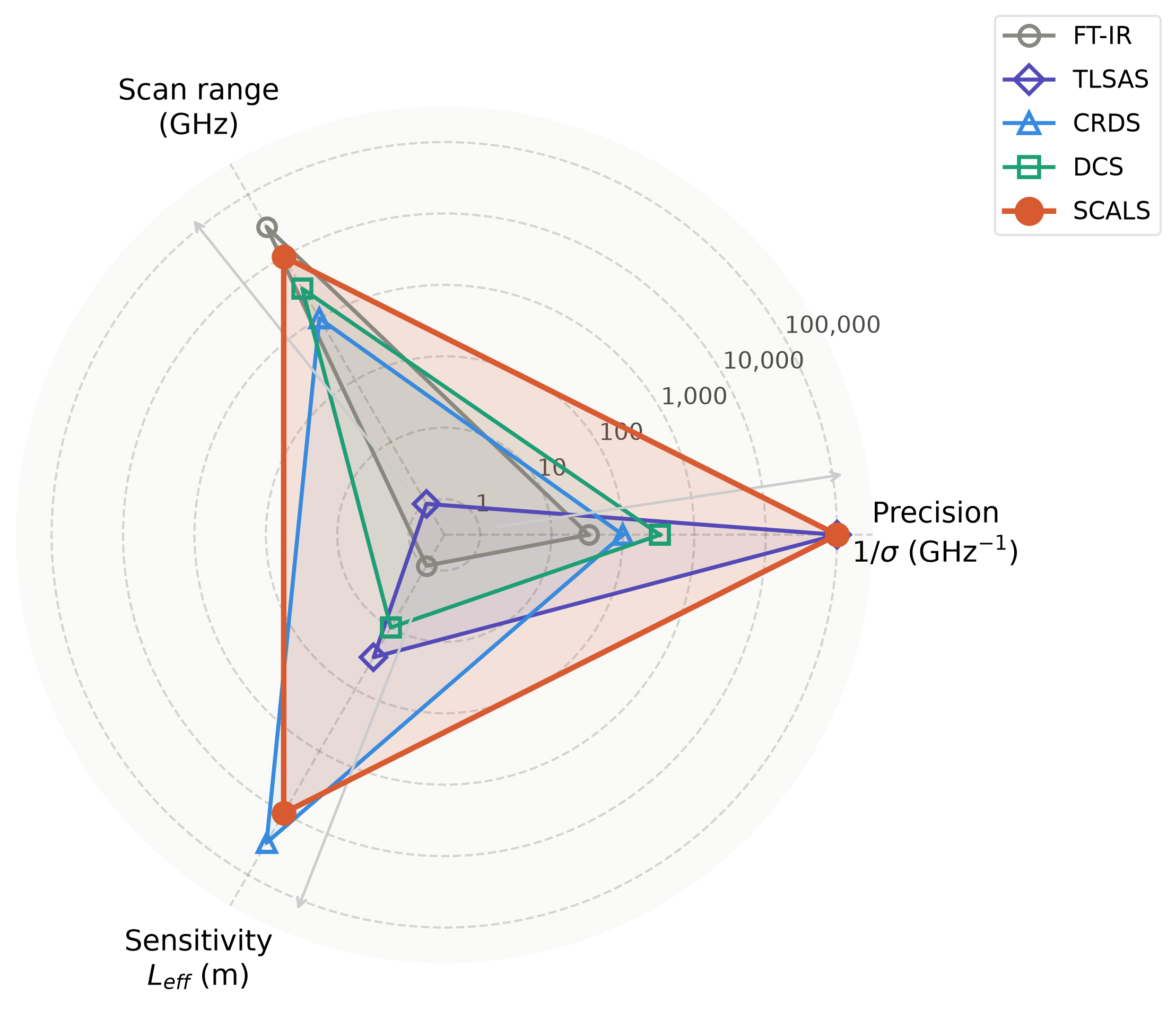}
  \caption{Multi-parameter comparison of spectroscopic techniques: Fourier-transform infrared spectroscopy (FT-IR), tunable laser saturated absorption spectroscopy (TLSAS), cavity ring-down spectroscopy (CRDS), dual-comb spectroscopy (DCS), and Sequential Cavity-mode Auto-Lock Spectroscopy (SCALS, this work). Three axes (scan range, precision ($1/\sigma$), and sensitivity ($L_{\mathrm{eff}}$)) are plotted on a $\log_{10}$ absolute scale, with precision shown as its reciprocal so that all axes follow a uniform ``higher~$=$~better'' convention (outward). SCALS fills the previously unoccupied corner of the parameter space.}
  \label{fig:radar}
\end{figure}

Here we shift this paradigm to a purely data-driven one, introducing a model-free strategy that reconstructs molecular energy levels directly from the measured frequencies of unassigned transitions, using only the algebraic closure of the transition network, with no prior knowledge of quantum states and no Hamiltonian model. The essential requirement is a spectrum that is at once broad, densely sampled, measured with kilohertz accuracy, and sufficiently sensitive to resolve weak transitions. Satisfying these demands simultaneously has remained one of the central experimental challenges in molecular physics: the highest frequency accuracy demands slow, meticulous laser scanning, while rapid acquisition over broad bandwidths inevitably sacrifices precision~\cite{Riehle2018Met}, a limitation that becomes acute when searching for unknown transitions over extended ranges, as in the hunt for the nuclear clock transition in thorium-229~\cite{Tkalya1996PS-Th, Yamaguchi2024Nature-Th, Zhang2024Nature-Th}. As summarised in Figure~\ref{fig:radar}, Fourier-transform infrared spectroscopy (FT-IR)~\cite{Griffiths2007} offers the broadest scan range but limited precision and sensitivity; tunable laser saturated absorption spectroscopy (TLSAS)~\cite{Preston1996, Martin2016} gives excellent precision but a narrow scan range; cavity ring-down spectroscopy (CRDS)~\cite{Berden2000, Mazurenka2005} gives high sensitivity and precision but a restricted scan range; and dual-comb spectroscopy (DCS)~\cite{Picque2026, Picque2019NatPhoton, Coddington2016} covers a broad band with high precision but only moderate sensitivity without an enhanced cavity~\cite{Bernhardt2010NatPhoton, Hu2022ASR}. No existing technique simultaneously fills the corner of broad coverage, high precision, and high sensitivity.

We meet this requirement with a cavity-enhanced spectrometer, termed SCALS (Sequential Cavity-mode Auto-Lock Spectroscopy), which locks a tunable diode laser to a high-finesse optical cavity that serves simultaneously as the absorption cell, the frequency reference, and the scanning actuator. An automated cycle of fine cavity scan, auto-unlock and relock to the adjacent cavity mode, and coarse laser tuning steps the lock from one longitudinal mode to another, extending the total spectral coverage well beyond 10~THz with a single near-infrared diode laser. SCALS thus occupies the previously unoccupied corner of the precision-bandwidth-sensitivity parameter space shown in Figure~\ref{fig:radar}.

We demonstrate SCALS by measuring high-resolution spectra of acetylene ($^{12}$C$_2$H$_2$), methane ($^{12}$CH$_4$), and deuterated water mixture across a spectral range of $1.537- 1.691\;\mu$m, with a frequency sampling interval at the 0.1~MHz level and an absolute accuracy at the kilohertz level. These three molecular species span the archetypal categories of molecular spectroscopy: a linear polyatomic (C$_2$H$_2$) with a rich rovibrational structure that serves as a frequency reference in optical telecommunications~\cite{BIPM2024_C2H2, Madej2006JOSAB}, a spherical top (CH$_4$) with dense rotation-vibration-coupled bands of critical importance for atmospheric remote sensing missions such as MERLIN~\cite{Vasilchenko2023JQSRT}, and an asymmetric rotor (HDO), the simplest member of the largest and most structurally diverse class of molecules. Together they demonstrate the versatility of the technique across widely differing spectroscopic complexity.

These dense, broadband, kilohertz-accurate spectra then feed the model-free reconstruction introduced above. Together, SCALS and the model-free reconstruction provide a practical pathway to comprehensive, kilohertz-accurate spectral atlases covering tens of terahertz, with immediate implications for precision molecular physics, atmospheric science, and the spectroscopic characterisation of molecules of astrophysical, chemical, and metrological interest.

%%%%%%%%%%%%%%%%%%%%%%%%%%%%%%%%%%%%%
%\section{RESULTS}

\section*{SCALS: a broadband spectrometer for dense kilohertz-accurate spectra}

\begin{figure}
\centering\includegraphics[width=0.95\textwidth]{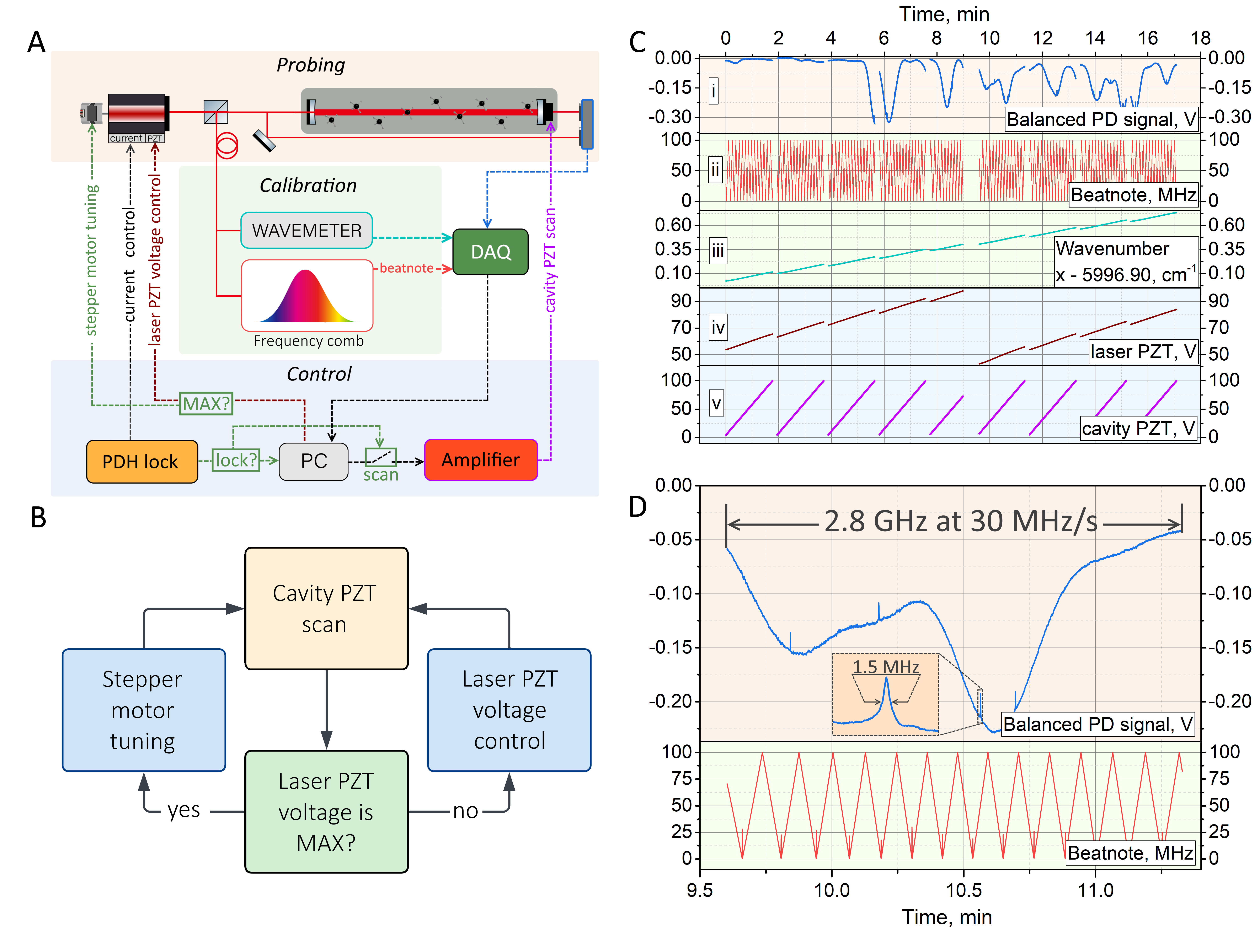}
\caption{Experimental setup and operational principle. 
(A) Schematic of the spectrometer. Wire colors correspond to signal colors in (C). Background colors indicate the three main functional subsystems: calibration (green), probing (beige), and control (blue). 
(B) Flowchart of the automated three-stage scanning procedure enabling continuous THz-range coverage. 
(C) Control and readout signals during a 22\,GHz automated scan (17 minutes duration): (i) cavity transmission spectrum; (ii) beat-note frequency between probe laser and optical frequency comb; (iii) approximate wavenumber from wavelength meter; (iv) laser PZT voltage controlled by the digital PID loop; (v) cavity PZT voltage ramped to scan the frequency. The laser current (not shown) is simultaneously controlled by the fast analog PID loop. 
(D) Sample spectrum of CH$_4$ at 1667.125~nm, showing saturated absorption features (Lamb dips) resolved on Doppler-broadened profiles. Panels A, C, and D are color-coded according to subsystem functionality.
    \label{fig:setup}}
\end{figure}

SCALS produces these dense, broadband, kilohertz-accurate spectra by scanning continuously across tens of terahertz. Broad coverage is achieved through a three-stage scanning process: 
(i) fine continuous scanning by piezoelectric transducer (PZT) actuation of the cavity mirror, providing $\sim 3$~GHz (0.1 cm$^{-1}$) per scan while maintaining the laser-cavity lock; 
(ii) intermediate-range laser frequency tuning via the laser's internal PZT, enabling mode-hop-free scans of approximately 15 GHz; and 
(iii) coarse grating adjustment using a stepper motor, extending the total scan range beyond 1 THz. The control sequence is illustrated in Figs.~\ref{fig:setup}A and~\ref{fig:setup}B. Figure~\ref{fig:setup}C shows the control and reading signals acquired during a continuous 22 GHz scan performed automatically within 17 minutes, while Fig.~\ref{fig:setup}D presents a calibrated spectrum of CH$_4$ at 1667 nm, where saturated absorption lines (Lamb dips) are clearly resolved on top of Doppler-broadened features.

Absolute frequency calibration is provided by simultaneous monitoring with an optical frequency comb (via beat-note detection) and a high-precision wavelength meter. A custom control system integrates wavelength meter readings with beat-note tracking to dynamically manage lock acquisition, mode hopping, and scan sequencing without operator intervention. This sequential cavity-mode auto-lock strategy enables unattended, continuous spectral acquisition across the full tuning range.

Beyond coverage and precision, the high-finesse cavity supplies the third requirement: sensitivity. Equipped with two mirrors with $R\simeq 0.99997$, the 42-cm-long cavity folds the absorption into an effective path length of approximately 28~km, and the noise level is further reduced by the balanced detection and laser power-stabilizing scheme, so that even the weak, closely spaced transitions of polyatomic molecules are recorded with high signal-to-noise ratio. 

The hallmark of the SCALS approach is its ability to perform cyclic lock-scan-unlock-relock sequences at high speed, enabling seamless coverage over a broad spectral range. Maintaining the laser-cavity lock during rapid scanning requires real-time coordination between multiple frequency references and control loops. A dual-loop Pound-Drever-Hall (PDH) scheme ensures continuous resonance: a fast analog loop modulates the laser current for tight locking, while a slower digital loop adjusts the laser PZT voltage to track the shifting cavity mode. This configuration enables instantaneous scan rates up to 30 MHz/s, two orders of magnitude faster than conventional cavity-enhanced systems~\cite{Wang2017JCP}. Critically, the digital loop also manages the sequential cavity-mode auto-locking procedure: after each scan, it deliberately unlocks the laser, sweeps the laser PZT to find the next cavity mode, and relocks to that mode, all using real-time feedback from a wavelength meter and a beat-note signal against an optical frequency comb.

\begin{figure}
\centering\includegraphics[width=0.8\textwidth]{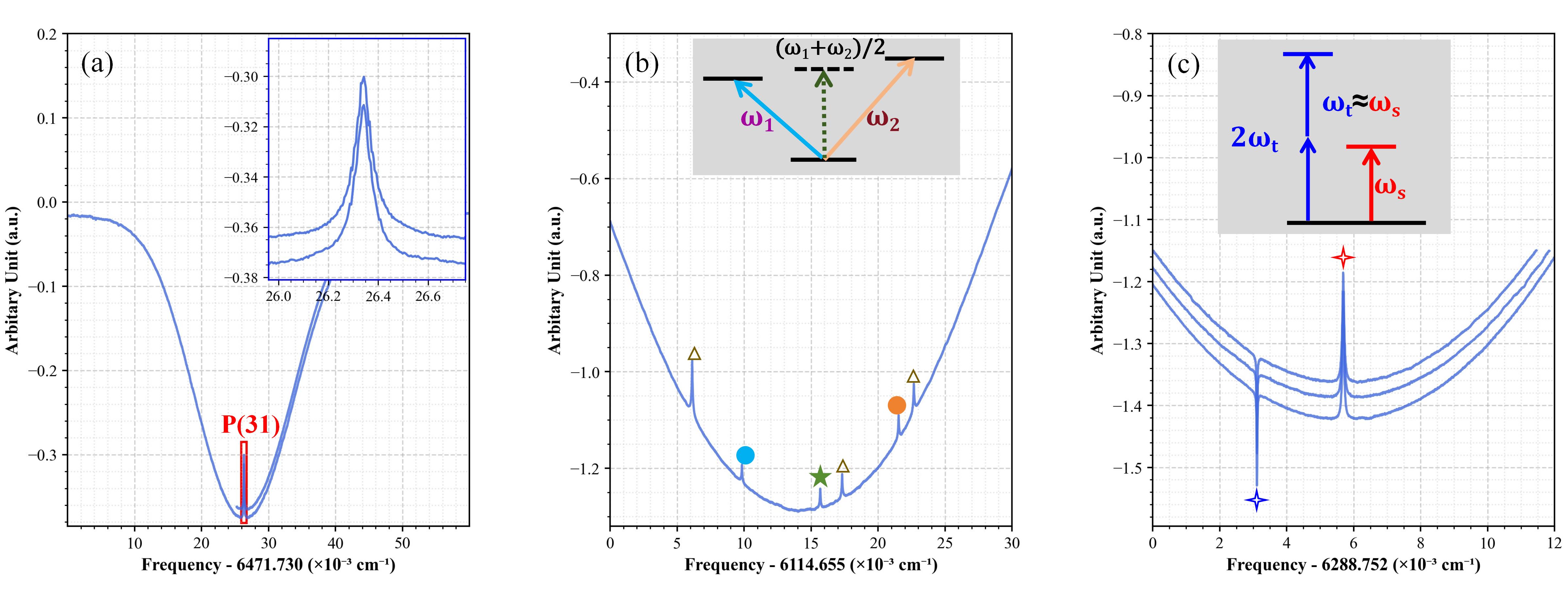}
\caption{Sample spectra recorded by SCALS.
(a) The P(31) line of $^{12}$C$_2$H$_2$ at 6471.730~cm$^{-1}$. Spectra recorded in two different scans are illustrated.
(b) CH$_4$ spectrum near 6114.655\,cm$^{-1}$. The peak marked with a star is a crossover line of two nearby lines indicated with blue and red dots.
(c) Repeated scan of HDO spectrum near 6288.752\,cm$^{-1}$. The dip at 6288.7551\,cm$^{-1}$, different from the conventional SAS peak (6288.7577\,cm$^{-1}$), corresponds to a two-photon absorption line.
    \label{fig:sample-spec}}
\end{figure}

The effective spectral acquisition rate is roughly 2~cm$^{-1}$ per hour, spanning hundreds of thousands of spectral points. This throughput is constrained by necessary overhead between scans, including stepper-motor coarse tuning, laser parameter optimization for single-mode operation, cavity mode searching, and relocking procedures, all managed automatically by the control system. Despite this limitation, the system maintains frequency precision at the kilohertz level across its entire tuning range.
The measurement precision is ultimately limited by the scan speed and sampling step size. The observed saturated absorption linewidths are currently limited by transit-time broadening of a few hundred kilohertz, and the scan speeds and step sizes employed in this work are matched to this linewidth. For transitions requiring narrower linewidths or higher precision, the sampling step size can be reduced and the scan speed decreased accordingly.
Detailed experimental procedures are provided in the \textit{Methods} section, and Figs.~\ref{fig:sample-spec}(a-c) present sample spectra of C$_2$H$_2$, CH$_4$, and HDO, respectively.

%%%%%%%%%%%%%%%%%%%%%%%%%%%%%%%%%%%

%%%%%%%%%%%%%%%%%%%%%%%%%%%%%%%%%%%

%\subsection*{Precision benchmark: acetylene lines against frequency references}

\textit{Precision benchmark.} To validate the precision and repeatability of our method, we performed multiple measurements of the Lamb dip of the P(31) line in the $\nu_1+\nu_3$ band of $^{12}$C$_2$H$_2$ under varied experimental conditions (scan rates, directions, input powers, and acquisition synchronization). A portion of the spectrum, recorded in two different scans, is shown in Fig.~\ref{fig:sample-spec}(a). The measured line positions exhibited a power-dependent shift of approximately $-12$ kHz/mW, with single-measurement standard deviations of 10--30 kHz. Independent calibration using four consecutive beat-note readings per spectral point yielded line positions consistent within 17 kHz across all conditions, with no systematic trend. At the scan rate of 21 MHz/s, the frequency coverage during the 5 ms integration time is only 105 kHz, confirming adequate synchronization. All results are in agreement with the known transition frequency of the P(31) line.
As an extended validation, we continuously recorded the spectrum of C$_{2}$H$_2$ in the 6471--6505~cm$^{-1}$ range, obtaining Lamb dips for 129 transitions. Of these, 18 transitions correspond directly to Lamb-dip positions available in literature~\cite{BIPM2024_C2H2, Edwards2005, Castrillo2023}, with deviations in the range between $-180$~kHz and $80$~kHz (Fig.~\ref{fig:C2H2-SAS} in \textit{Supplementary Information}), confirming the measurement accuracy of our instrument. The 129 Lamb dips detected span line intensities from $1.4\times10^{-21}$ down to $3.0\times10^{-24}$~cm/molecule, a range of almost three orders of magnitude, confirming that the spectrometer resolves transitions far weaker than typical reference lines while retaining kilohertz-level frequency precision.
Detailed experimental conditions and data are provided in \textit{Supplementary Information}.

%\subsubsection*{Crossover Resonances}

\textit{Crossover resonances.} In saturation spectroscopy, spurious dips (crossover resonances) occur due to the interaction between two closely spaced molecular transitions $(\omega_1, \omega_2)$ sharing a common lower (upper) energy level. Molecules with non-zero longitudinal velocity $v_z$ experience simultaneous saturation of both transitions through the Doppler frequency shift effect, resulting in an absorption dip at the central frequency $\omega_0 = (\omega_1 + \omega_2)/2$.
Since a crossover resonance only occurs when two transitions share a common upper ($\Lambda$-type) or lower (V-type) energy level, it provides a useful diagnostic for line assignment in densely congested spectra. 
As shown in Fig.\,\ref{fig:sample-spec}b, we observed a crossover line of CH$_4$ at 6114.6707~cm$^{-1}$, which was misinterpreted as a genuine molecular transition in the work by Votava et al.\,\cite{Votava2022} (labeled R10-8 in their Table 1), whereas the actual transition at $6114.664840~\text{cm}^{-1}$ went undetected. This error highlights a fundamental limitation of selective spectral scanning near predicted line positions: when theoretical predictions are inaccurate, an incomplete spectral scan cannot distinguish artifacts from genuine transitions. 
Continuous recording over the full investigated range, as practised here, avoids such errors by detecting all spectral features and reliably identifying cross-saturation dips.

%\subsubsection*{Two-photon absorption (TPA) lines}
\textit{Two-photon transitions.} The present broadband scanning capability provides a versatile method to search for two-photon absorption (TPA) lines. Molecular TPA transitions are hard to predict and present a significant technical challenge, as each requires a single-photon absorption (SPA) transition to a real intermediate state adjacent to the virtual intermediate state of TPA, and the TPA cross-section is very sensitive to the distance between the real and virtual intermediate states. Two-photon transitions are attractive as frequency standards because their first-order Doppler shift cancels when the counter-propagating photons have the same frequency. TPA transitions also prove particularly valuable for trace gas detection, combining enhanced spectral selectivity compared to Doppler-limited spectroscopy with the prospect of achieving both high sensitivity and high selectivity \cite{Lehmann2019JCP, Zhao2020PRA-2P}. Recently, we demonstrated the detection of trace radioactive $^{14}$CO$_2$ through TPA spectroscopy~\cite{Liu2025AC, Tan2026SenAB}.
In the wide spectral range we covered in this work, we found one TPA transition for $^{12}$C$_2$H$_2$ and one for HD$^{16}$O. Both are identified for the first time. 
The $^{12}$C$_2$H$_2$ TPA line is located at 6471.920943~cm$^{-1}$, enhanced by the nearby SPA transition R(9e) [$\nu_2+\nu_3+2\nu_4^0~\Sigma_u^+$] at 6471.921716~cm$^{-1}$, which is also observed as a Lamb dip.
A TPA transition of HD$^{16}$O is found at 6288.755116~cm$^{-1}$, as shown in Fig.\,\ref{fig:sample-spec}c. It can be assigned as originating from the $3_{3~1}$ [0~0~0] level in the ground state to the upper state of $3_{2~1}$ [1~2~3] (noted as $J_{K_a K_c}$ [$v_1, v_2, v_3$]). The TPA transition is enhanced by the SPA transition at 6288.757691~cm$^{-1}$ to the intermediate state of $2_{2~0}$ [1~0~1]. The TPA line yields an upper-level energy of 12810.534032~cm$^{-1}$, which is 0.0031~cm$^{-1}$ higher than the MARVEL value of 12810.5309~cm$^{-1}$~\cite{Furtenbacher_MARVEL_JMS}.

%%%%%%%%%%%%%%%%%%%%%%%%%%%%%%%%%
\section*{Model-free reconstruction of molecular energy levels by inverse graph construction}

Traditional analysis is model-driven: methods such as AUTOFIT~\cite{Seifert_Autofit} fit and refine an effective Hamiltonian against the data, and can fail under severe quantum perturbations. Reconstructing energy levels from unlabeled spectra alone is a long-standing challenge, but it is an instance of a more general problem, ``topological discovery'', that has been solved in other fields.

In bioinformatics, de novo genome assembly stitches together gene fragments from de Bruijn and overlap graphs without a reference template~\cite{Bankevich_genome, Kamath_Genome}; in network science, topology inference reconstructs hidden graphs from nodal signals~\cite{Segarra_Network, Dong_Network}; and in computer vision, structure-from-motion recovers camera poses from image correspondences~\cite{Schonberger_SFM}. In each case, a latent network is reconstructed from relative measurements alone, by enforcing algebraic loop closures (such as K$_{2,~2}$ bipartite cycles) and performing global synchronization over the Graph Laplacian~\cite{Cucuringu_Laplacian, Jiang_Laplacian}.

Molecular spectroscopy can exploit the same principle. When transition frequencies are measured to the $3\times 10^{-6}$ cm$^{-1}$ precision achieved here, energy conservation becomes a rigorous algebraic invariant: every closed cycle of transitions must sum to zero within the measurement uncertainty, so the energy-level topology can be recovered from the data alone, independent of any empirical model.

Given only a list of high-precision, unassigned transition frequencies $F=\{f_1, f_2, \ldots, f_N\}$, the algorithm reconstructs the underlying bipartite energy-level network, composed of the upper-state set $U=\{E^\text{U}_1, E^\text{U}_2, \ldots\}$ and the lower-state set $L=\{E^\text{L}_1, E^\text{L}_2, \ldots\}$, in which every transition $f_k$ corresponds to an edge of weight $f_k = E^\text{U}_n - E^\text{L}_m$, via the following inverse-graph-construction (IGC) procedure.

\begin{figure}
    \centering
    \includegraphics[width=0.9\textwidth]{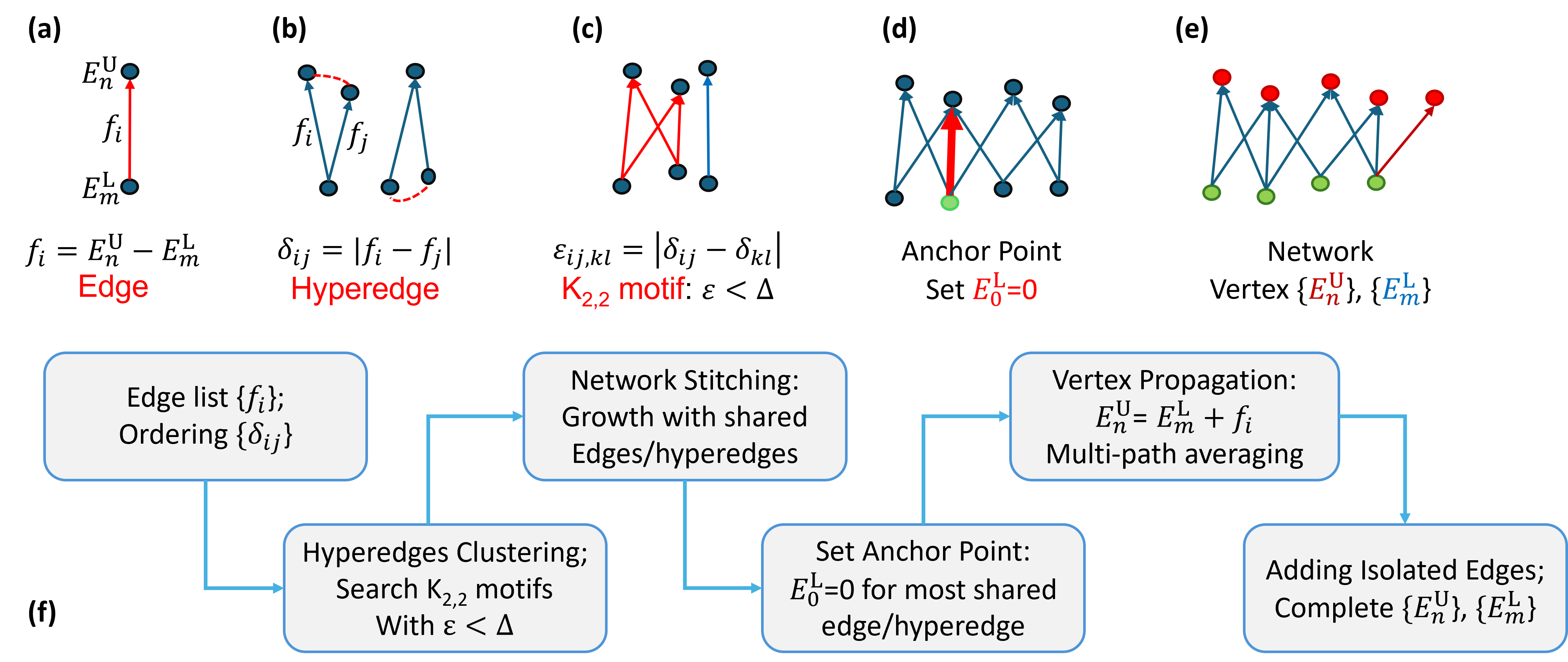} 
    \caption{\textbf{Schematic of the inverse graph construction (IGC) algorithm.} 
    (a) Unassigned high-precision transition list as edge set $\{f_{i}\}$, with each $f_i$ connecting two vertices (upper energy level $E^U_n$ and lower energy level $E^L_m$); 
    (b) Difference clustering identifies candidate hyperedges (recurring combination differences) $\{\delta_{ij}\}$. 
    (c) 4-cycle ($K_{2,2}$) motif detection with preset threshold of $\varepsilon = |\delta_{ij}-\delta_{kl}|<\Delta$. 
    (d) Network stitching via union-find merges fragments into connected components. Set the most shared edge as an anchor, letting the lower energy be $E^L_0 = 0$;
    (e) Breadth-first propagation yields the reconstructed energy-level network.
    (f) Summary flowchart.}
    \label{fig:ARalgorithm}
\end{figure}

\textbf{(1) Mining candidate hyperedges by difference clustering.}
All pairwise differences, $\delta_{ij} = |f_i - f_j|$, are clustered in one dimension within the tolerance of $\Delta$ (e.g., sliding-window grouping). Difference clusters with occurrence count $\ge 2$ are extracted as hyperedges; physically, each recurring difference represents an upper- or lower-state energy combination difference, and topologically it links the edges that share a common vertex.

\textbf{(2) Capturing 4-cycles (instantiating local vertices).}
A difference alone cannot distinguish an upper-state from a lower-state gap, but two equal differences close the graph. The algorithm searches for quadruplets $(f_i, f_j, f_k, f_l)$ satisfying the combination condition $|\delta_{ij} - \delta_{kl}| < \Delta$. Each such quadruplet inevitably forms a complete bipartite subgraph K$_{2,2}$ motif shown in Fig.~\ref{fig:ARalgorithm}c: four local vertices are instantiated, recording only their topological connectivity for the moment.

\textbf{(3) Network stitching and growth.}
The independent 4-cycle fragments are merged: cycles sharing edges or hyperedges share vertices, so a union-find (disjoint set) procedure stitches the fragments into several connected components of the bipartite graph.

\textbf{(4) Boundary condition and vertex assignment.}
Within the largest connected component, an anchor edge $f_{\mathrm{start}}$ is selected, and the numerical relativity is broken by setting the reference level $E^{\text{L}}_0 = 0$ and $E^{\text{U}}_0 = f_{\mathrm{start}}$. A breadth-first search then propagates the values along the topology: from a known $E^{\text{L}}_m$, every edge $f_i$ yields $E^{\text{U}}_x = E^{\text{L}}_m +f_i$; from a known $E^{\text{U}}_n$, every edge $f_j$ yields $E^{\text{L}}_y = E^{\text{U}}_n - f_j$. Because closed cycles give multiple routes to the same vertex, the slightly discrepant values (due to the tolerance $\Delta \simeq 5\times10^{-6}$ cm$^{-1}$ related to the experimental uncertainty) are averaged to obtain the initial energy level.

\textbf{(5) Attaching isolated edges.}
Frequencies that did not participate in any 4-cycle K$_{2,2}$ motif are matched against the established sets $U$ and $L$; if $f_i \approx E^{\text{U}}_\text{new} - E^{\text{L}}_m$ with $E^{\text{U}}_{\text{new}}$ in a physically reasonable range, a new upper-state vertex is created and attached to the network.

The algorithm outputs the upper- and lower-state energy-level sets $U$ and $L$, thereby completing the inverse reconstruction of the quantum energy-level network from unassigned high-precision spectra. In essence, \textit{differences reveal connections (hyperedges)} $\rightarrow$ \textit{4-cycles establish bipartite local subgraphs K$_{2,2}$} $\rightarrow$ \textit{cycle merging generates the global topology} $\rightarrow$ \textit{setting a zero-point enables network-wide assignment}.  
Governed by the strict parity selection rule for electric dipole transitions, the upper and lower energy levels of an absorption spectrum naturally constitute two disjoint vertex sets (upper {$U^{\pm}$} and lower {$L^{\mp}$}) of a bipartite graph. 

\begin{figure}
\centering
\includegraphics[width=0.9\textwidth]{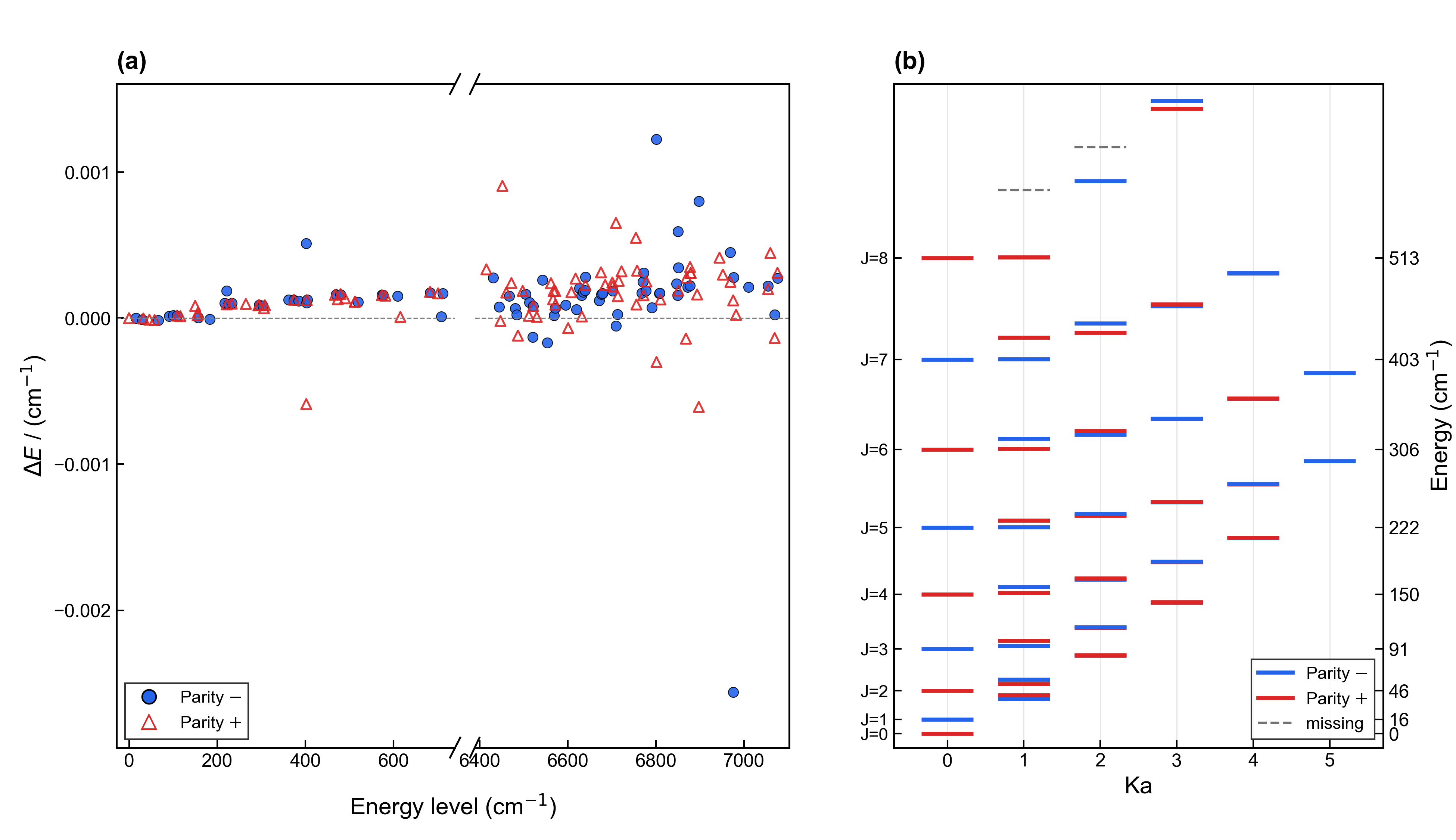} 
\caption{\textbf{Energy levels reconstructed by a data-driven algorithm.} 
    (a) Differences between HDO energies obtained by IGC in this work and MARVEL values~\cite{Tennyson2010}, $\Delta E = E_\textrm{MARVEL} - E_\textrm{this work}$, reflecting the larger uncertainties of the MARVEL values.
    (b) Rotational quantum numbers of ground-state energy levels, assigned a posteriori by comparison with MARVEL. Dashed lines indicate missing levels due to transitions beyond the laser tuning range.
    }
    \label{fig:HDO-dE}
\end{figure}

We applied this graph-theoretic approach to the high-precision absorption spectrum of HD$^{16}$O, reconstructing 158 energy levels, grouped into two sub-networks, from 686 unassigned transitions.
The reconstructed levels are determined at kilohertz-level precision, exceeding that of the MARVEL database of HDO energies~\cite{Tennyson2010}, as shown in Figure~\ref{fig:HDO-dE}(a). Because the MARVEL levels carry well-established quantum-number labels, this comparison serves only to assign the labels a \textit{posteriori}; the reconstruction itself requires none.
A handful of levels, however, deviate from MARVEL beyond its stated uncertainties, and each such outlier is traced to a level whose energy in MARVEL was derived from blended, Doppler-broadened spectra, which the kilohertz resolution of SCALS resolves cleanly (see Supplementary Section S3).
Figure~\ref{fig:HDO-dE}(b) shows the a posteriori quantum-number assignment of the reconstructed ground-state levels, with the missing high-$J$ levels indicated as dashed lines.

The key to this reconstruction is that measurement precision acts as a topological criterion: because a random K$_{2,2}$ (4-cycle) closure within tolerance $\Delta$ is vanishingly improbable, a recurring frequency difference is near-deterministic evidence of genuine connectivity, enabling fully automated inference of energy-level networks from unassigned spectra.

%%%%%%%%%%%%%%%%%%%%%%%%%%%%%%%%%%%%%%%%%%%%

\section*{DISCUSSION}

The central result of this work is a purely data-driven, model-free reconstruction of molecular energy levels directly from unassigned transitions. This reconstruction is enabled by a broadband, kilohertz-accurate spectrometer (SCALS), whose densely sampled spectra span tens of terahertz while maintaining frequency uncertainties at the tens of kilohertz level.

The success of this reconstruction rests on the three measurement qualities introduced at the outset. Sufficient sensitivity and spectral coverage are required to detect enough transitions for a well-connected, high-confidence energy-level network; without them, levels whose connecting transitions are too weak, or that fall outside the scanned range, are simply absent from the network, and the reconstruction is left with isolated fragments rather than a complete connected component. Sufficient precision is required for the opposite reason: it reduces the probability of accidental closures among the frequency differences, so that a K$_{2,2}$ motif in the data can be trusted as genuine connectivity rather than coincidence. This is what allows true motifs to be identified and the energy-level network to be evolved reliably even within a dense forest of transitions. Sensitivity, coverage, and precision are therefore not independent design goals but joint prerequisites for model-free reconstruction.
The instrument's high-speed, high-precision, and high-sensitivity capabilities directly enable broadband detection of nonlinear spectral features, as demonstrated by the unambiguous observation of Lamb dips, crossover resonances, and two-photon absorption spectral signs. This capacity to resolve complex, weak nonlinear interactions across wide spectral ranges addresses a critical limitation in existing techniques, which typically require prior knowledge of transition locations or sacrifice bandwidth for precision.

The methodology establishes a foundation for future instrumentation capable of supporting both exploratory spectroscopic studies and metrology-grade measurements where broad bandwidth and high resolution are required simultaneously. For example, the reconstructed energy-level differences span microwave-to-terahertz frequencies, providing kilohertz-accurate rest frequencies to guide precision radio astronomy. Such advancements would be particularly valuable for the development of the spectroscopic networks approach \cite{Csaszar2016, Tobias2020} and enhancing the accuracy and completeness of spectroscopic databases.
We also demonstrated that broadband precision spectroscopy enables detection of narrow (and usually also weak) transitions without any prior knowledge. Such transitions could help to address several fundamental physics problems \cite{Safronova2018RMP} by systematically unlocking elusive, highly sensitive transitions to probe extremely weak effects, such as the time variation of fundamental constants. The data-driven reconstruction of energy levels demonstrated in this work opens the door to systematic studies of polyatomic molecules whose complex intramolecular interactions have long hindered high-resolution spectroscopy.

%\section{Materials and Methods}
\section*{Methods}

%\subsection{SCALS principle and experimental apparatus}

The SCALS method is built around a high-finesse optical cavity that serves both as a frequency reference and as a scanning actuator. The light source is an external-cavity diode laser (ECDL, TOPTICA DL PRO) tunable over the near-infrared. Its frequency is phase-modulated by an electro-optic modulator (EOM) and locked to the cavity using the PDH technique. The optical cavity is housed in vacuum and consists of two mirrors with reflectivity $R \approx 0.99997$, separated by 42 cm, yielding a finesse of approximately $10^5$. One mirror is mounted on a piezoelectric transducer (PZT); applying a voltage ramp from 0 to 100 V continuously changes the cavity length and, consequently, the laser frequency over 3 GHz (0.1 cm$^{-1}$).

A dual-loop PDH system maintains lock during scanning. The fast analog loop modulates the laser injection current (``current control'' in Fig.~\ref{fig:setup}), providing tight locking within a narrow range. The slow digital loop (through a LabVIEW program) adjusts the laser PZT voltage (``voltage control'') to track the cavity resonance during extended scans. This digital loop also handles mode selection and automatic relocking based on feedback from the wavelength meter and beat-note signal.

%\subsection{Sequential cavity-mode auto-locking procedure}

Automated scanning across tens of terahertz follows a three-step sequence (Fig.~\ref{fig:setup}B), orchestrated by a custom control system that integrates real-time wavelength meter readings and beat-note tracking. The entire procedure is driven by repeated sequential cavity-mode auto-lock cycles:

\begin{enumerate}
    \item \textbf{Fine cavity PZT scan.} A linear voltage ramp applied to the cavity PZT shifts the cavity modes (``frequency scan'' in Fig.~\ref{fig:setup}A). The dual PID loops keep the laser locked during the ramp, covering $\approx 0.1$ cm$^{-1}$ (3 GHz) per scan. The wavelength meter continuously verifies the scan progression.
    
    \item \textbf{Auto-unlock and relock to the next mode.} At the end of each ramp, the cavity PZT voltage is reset to zero while the laser frequency is held constant. The control system then deliberately unlocks the laser, sweeps the laser PZT to locate the nearest cavity mode, and auto-relocks to that mode, all without user intervention. This lock-scan-unlock-relock cycle repeats after every single scan, sequentially stepping the locking point from one cavity mode to the next.

    \item \textbf{Coarse laser tuning.} After several such cycles, when the laser's internal PZT reaches its voltage limit for mode-hop-free operation (approximately 15 GHz cumulative range), it is reset to its minimum. The control system then engages a stepper motor attached to the ECDL grating (``coarse tuning'' in Fig.~\ref{fig:setup}A), advancing the laser frequency by roughly 15 GHz per step. The sequential auto-lock routine then resumes, enabling continuous coverage of over 1~THz.
\end{enumerate}

This sequential cavity-mode auto-lock strategy allows unattended operation over extended periods, with the system repeatedly locking, scanning, unlocking, auto-relocking to successive cavity modes, and continuing without manual intervention while maintaining kilohertz-level frequency accuracy throughout the entire scan range.

Spectra are acquired with a 5 mV step on the cavity PZT, corresponding to a frequency spacing of 160 kHz between points. The integration time per point is 5 ms, with all data synchronously tagged with wavelength meter and beat-note readings.

%\subsection{Frequency calibration}

Absolute frequency calibration is performed using an optical frequency comb (OFC) referenced to a GPS-disciplined rubidium clock. The OFC is generated by an Er-doped fiber oscillator with repetition frequency $f_r \approx 200$ MHz and carrier-envelope offset frequency $f_0 = 250$ MHz. The OFC fractional frequency accuracy exceeds $1 \times 10^{-12}$, corresponding to 0.19 kHz at 1.6 $\mu$m.
The beat note between the probe laser and the nearest comb mode is detected by a fast photodiode (ET-3000A, Electro-Optics), amplified, filtered (100 MHz low-pass), and acquired by a high-speed digitizer (PXIe-5160, National Instruments). Beat-note frequencies in the 2--100 MHz range are recorded synchronously with spectral data. A high-precision wavelength meter (HighFinesse WS7) provides continuous frequency monitoring at a 1 Hz update rate, serving as both a real-time guide for the control system and a redundant frequency reference. The standard deviation of individual beat-note measurements is approximately 100 kHz, enabling kilohertz-level frequency precision after averaging.

%\subsection{Baseline stabilization}

Baseline fluctuations are suppressed by two complementary methods. First, a balanced photodetector (Thorlabs PDB450C) at the cavity output cancels common-mode laser amplitude noise. Second, the input laser power is actively stabilized via a feedback loop that adjusts the radio-frequency power driving an acousto-optic modulator (AOM), ensuring constant power coupled into the cavity.

%\subsection{System control and synchronization}

All components are coordinated through custom LabVIEW software that enables simultaneous data acquisition, laser control, and real-time monitoring. The control system continuously processes wavelength meter readings and beat-note signals to manage the lock-scan-unlock-relock cycles, automatically adjusting scan parameters and initiating relocking sequences when deviations are detected. This integrated approach ensures precise synchronization of tuning, locking, and frequency calibration throughout automated, long-duration measurements without operator intervention.

Spectra of C$_2$H$_2$, CH$_4$, and HDO were recorded at room temperature (296~K). Natural acetylene and methane samples with purity over 99\% were used, and the sample pressures were from 0.3 to 1 Pa. In HDO measurement, a 1:1 mixture of water and heavy water (D$_2$O, $>90$\% purity) with a total sample pressure of 1.5 Pa was used. The H$_2$O/HDO/D$_2$O ratio is estimated to be 1:2:1 in the sample.

\section*{Acknowledgments}
This work was jointly supported by  
 the National Natural Science Foundation of China (Grant Nos. 22273096, 22327801, 12393825, and 42550124), 
 the Independent Deployment Project of HFNL (Grant No. ZB2025010400), % length metrology
 the Quantum Science and Technology -- National Science and Technology Major Project (Grant Nos. 2021ZD0303102, 2021ZD0303304, and 2023ZD0301000), 
 and the Chinese Academy of Sciences (XDB0970100, XDA0520304).

\section*{Author contributions statement}

S. Vasilchenko: Methodology, Data Curation, Writing Original Draft;
A.-W. Liu: Methodology, Data Curation, Conceptualization, Writing --- Original Draft \& Editing;
C.-X. Zuo: Investigation, Data Curation;
Y.-Q. Cheng: Investigation, Data Curation;
Z.-T. Zhang: Investigation, Formal Analysis;
W.-T. Wang: Investigation;
Y. R. Sun: Methodology;
Y. Tan: Validation, Formal Analysis;
S.-M. Hu: Conceptualization, Supervision, Project Administration, Writing --- Original Draft \& Editing.

\section*{Additional information}

The authors declare no competing interests.
 
\bibliographystyle{naturemag}
\bibliography{AutoScan}

%%%%%%%%%%%%%%%%%%%%%%%%%%%%%%%%%%%%%%%%%%%%%%%%%%%
%%%%%%%%%%%%%%%% START OF SUPPLEMENT %%%%%%%%%%%%%%%
\clearpage
%\nolinenumbers
%\spacing{1.2}
% Figures, tables, equations and pages in the supplement are numbered S1, S2 etc.
\renewcommand{\thefigure}{S\arabic{figure}}
\renewcommand{\thetable}{S\arabic{table}}
\renewcommand{\theequation}{S\arabic{equation}}
\renewcommand{\thesection}{S\arabic{section}}
\renewcommand{\thepage}{S\arabic{page}}
\setcounter{figure}{0}
\setcounter{table}{0}
\setcounter{equation}{0}
\setcounter{section}{0}
\setcounter{page}{1} % not 0 as \newpage already started a supplementary page
% References continue the numbering from the main text.

\textbf{\center\large{Supplementary Information for:\\
    Model-free Reconstruction of Molecular Energy Levels by Broadband Kilohertz-accurate Cavity-enhanced Spectroscopy
}}

\subsubsection*{Contents:}
\begin{itemize}
\item S1. Accuracy assessment with molecular frequency standards (C$_2$H$_2$ P(31) validation)
\item S2. Broadband survey spectra for C$_2$H$_2$ and HDO (comparing with HITRAN database)
\item S3. Inverse graph construction of HDO energy levels
\end{itemize}

\clearpage 

\section{Accuracy assessment with molecular frequency standards}

To validate the precision and repeatability of our method, we performed multiple measurements of the Lamb dip of the P(31) line in the $\nu_1+\nu_3$ band of $^{12}$C$_2$H$_2$ under systematically varied experimental conditions, including different scanning rates (8, 13, and 21 MHz/s), opposite scanning directions, and varying input laser powers. We also assessed the synchronization between spectral data acquisition and frequency calibration. The number of measurements for each condition is summarized in Table~\ref{tab:scannumbers}.

Given the high scan rates employed, systematic shifts in the retrieved line positions could arise from either beat-note acquisition synchronization issues or a scan-rate-dependent response. We therefore analyzed these potential sources of systematic error using the collected data.

\begin{figure}
\centering\includegraphics[width=0.9\textwidth]{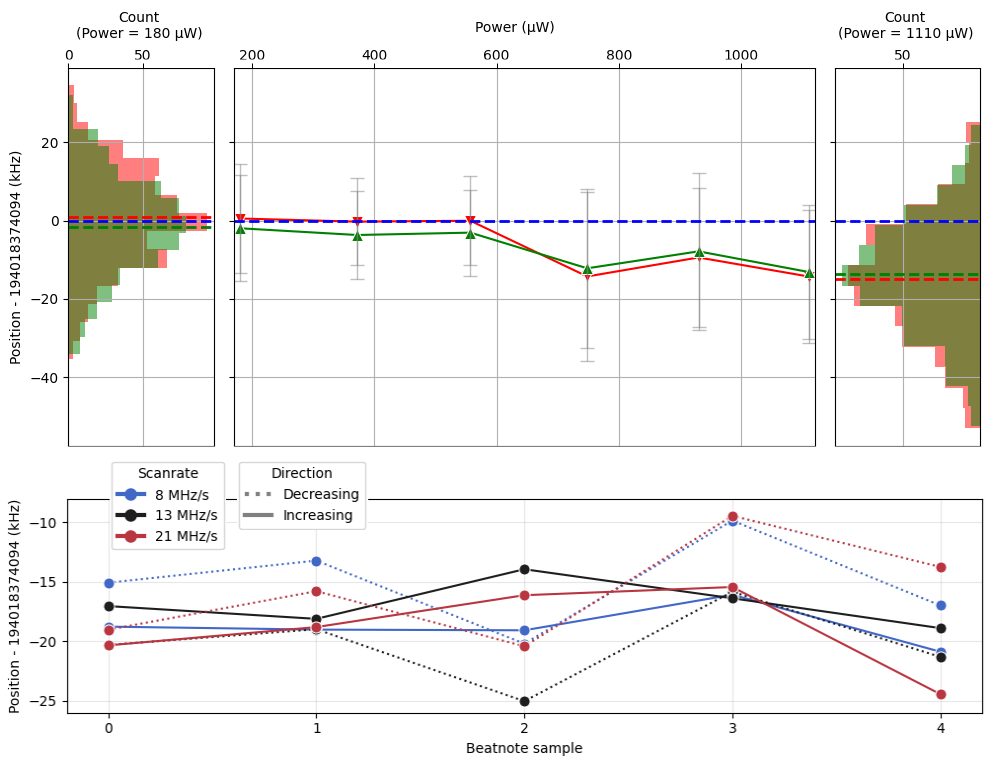}
\caption{\textit{Upper panel}: position of the P31 line obtained at different input laser power and scan directions at a scan rate of 21 MHz/s. Gray error bars indicate standard deviation of measurements. Histograms represent measurement distributions for scans towards higher (green) and lower (red) frequencies for the minimum and maximum power: 0.18~mW and 1.11~mW, respectively. The blue dashed line corresponds to the recommended value from Bureau Internationale des Poids et Mesures (BIPM)~\cite{BIPM2024_C2H2}. 
\textit{Lower panel}: Illustration on synchronization between spectral data and frequency reference acquisition (see text) for laser input power of 1.11 mW.
    \label{fig:position_vs_power_and_calibration}}
\end{figure}

The power dependence of the line position was measured at six injection power levels (upper panel of Fig.~\ref{fig:position_vs_power_and_calibration}). A linear fit yields a slope of approximately $-12$ kHz/mW, which is consistent with typical pressure- and power-shifting behavior observed in saturated absorption spectroscopy. The standard deviation of a single measurement ranges from 11 to 30 kHz, depending on the laser injection power, indicating good short-term repeatability.

To evaluate synchronization between spectrum acquisition and frequency calibration, we recorded four consecutive beat-note measurements during each 5 ms spectral point acquisition. Each spectrum was then independently calibrated using each of these four beat-note readings (labeled 1 to 4 in the lower panel of Fig.~\ref{fig:position_vs_power_and_calibration}, from first to fourth; label `0` denotes calibration using the averaged beat-note value). Regardless of the calibration method, scan rate, or scan direction, the resulting Lamb dip positions all fall within a range of approximately 17 kHz (lower panel of Fig.~\ref{fig:position_vs_power_and_calibration}), with no observable systematic trend. Notably, at a scan rate of 21 MHz/s, the system covers a frequency span of 17 kHz in just 1 ms, demonstrating that data acquisition synchronization remains satisfactory even at the highest scan speeds.

%%%%%%%%%%%%%%%%%%

All measured line positions are consistent with the known transition frequency of the $^{12}$C$_2$H$_2$ P(31) line, with deviations well within the combined statistical and systematic uncertainty. Given that the instrument's spectral step size is 160 kHz, the achieved resolution and accuracy at such rapid scanning rates represent a substantial advance in broadband precision spectroscopy. These results confirm that our method delivers high repeatability and accuracy across a wide range of experimental conditions without requiring prior knowledge of the target transition.

\begin{table}
    \centering
    \begin{tabular}{rrr}
        \hline
        Power & Decreasing & Increasing \\
        \hline
         0.180~mW & 549 & 544 \\
         0.372~mW & 562 & 562 \\
         0.557~mW & 565 & 565 \\
         0.748~mW & 513 & 504 \\
         0.930~mW & 502 & 502 \\
         1.110~mW & 599 & 599 \\
        \hline        
        Scan Rate & Decreasing & Increasing \\
        \hline
        8~MHz/s   & 431 &  432 \\
        13~MHz/s  & 377 &  377 \\
        21~MHz/s  &1815 & 1815 \\
        \hline
    \end{tabular}
    \caption{Number of measurements for different input laser powers (in mW), scan rates (in MHz/s), and directions. \label{tab:scannumbers}}
\end{table}

\begin{figure}
\centering\includegraphics[width=0.5\textwidth]{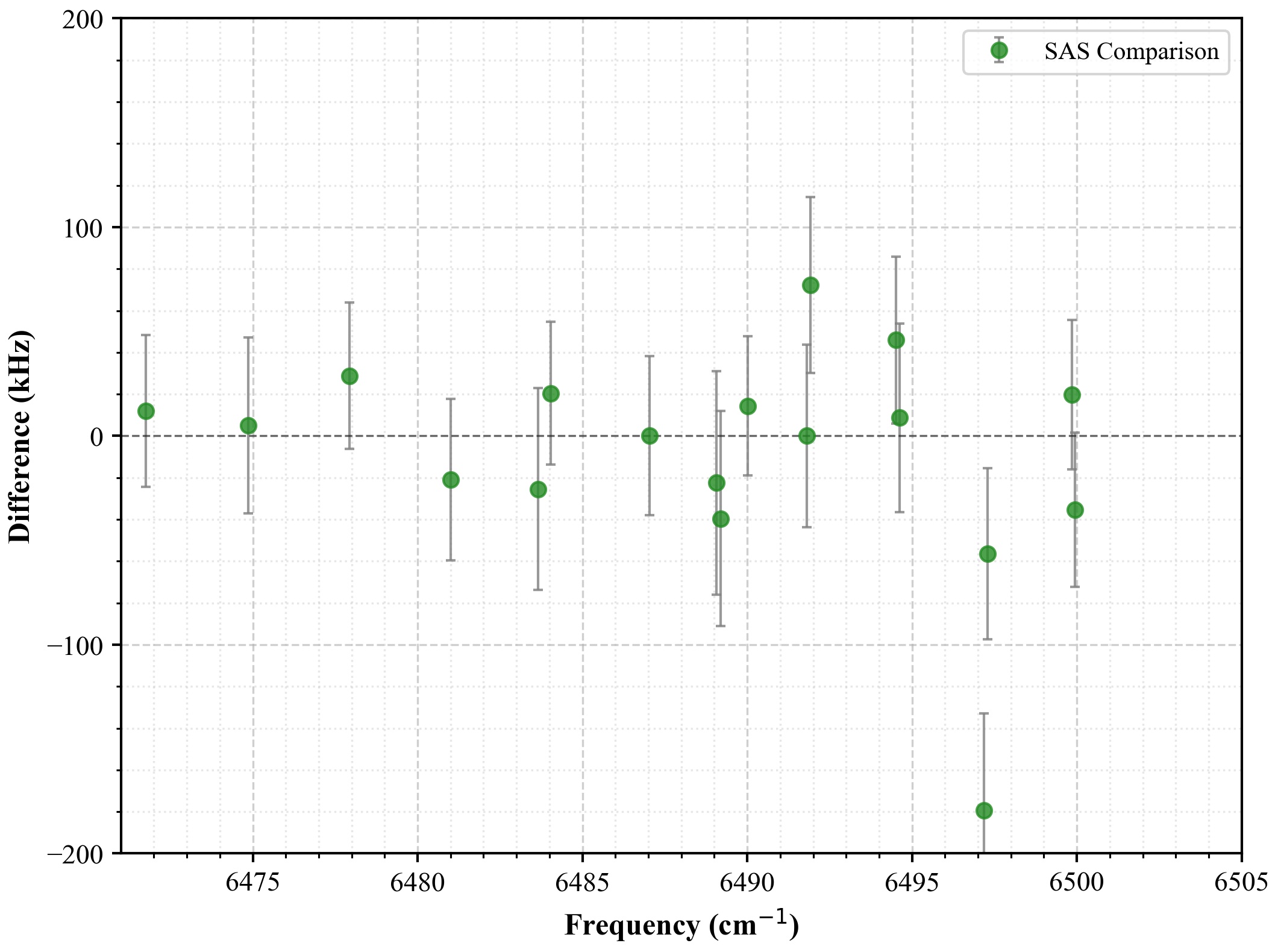}
\caption{Comparison of the C$_2$H$_2$ Lamb-dip positions obtained in this work and those from the literature~\cite{BIPM2024_C2H2,Edwards2005,Castrillo2023}.} 
    \label{fig:C2H2-SAS}
\end{figure}

We obtained 129 Lamb dips of C$_{2}$H$_2$ in the 6471--6505~cm$^{-1}$ range. %A full list, together with the assignments and shifts from the HITRAN database~\cite{GORDON2026HITRAN2024}, is given in the supplementary table. 
Among them, 18 transitions correspond directly to high-precision Lamb-dip positions available in the literature~\cite{BIPM2024_C2H2,Edwards2005,Castrillo2023}, and the differences are depicted in Fig.~\ref{fig:C2H2-SAS}. The deviations are in the range between $-180$~kHz and $80$~kHz, in agreement with the experimental uncertainty of this work.

%%%%%%%%%%%%%%%%%
\section{Line list and comparison with databases}

\begin{figure}
\centering\includegraphics[width=0.8\textwidth]{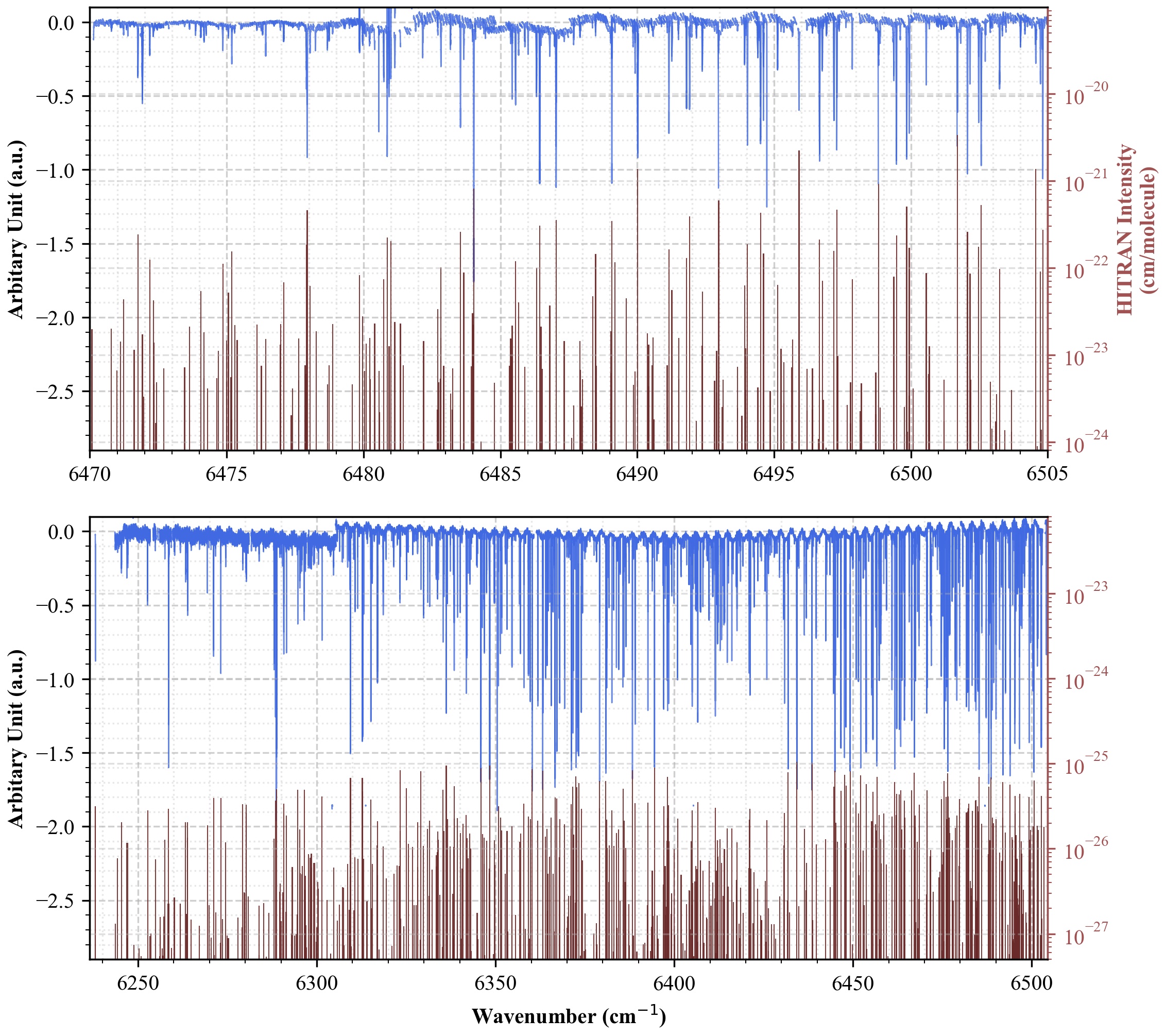}
\caption{Broadband spectra of C$_2$H$_2$ (upper panel) and HDO (lower panel) recorded by SCALS (blue traces). The lower brown bars indicate the HITRAN line lists.}
    \label{fig:overview-C2H2-HDO}
\end{figure}

Figure~\ref{fig:overview-C2H2-HDO} shows overviews of recorded SCALS spectra of C$_2$H$_2$ and HDO. For comparison, HITRAN~\cite{GORDON2026HITRAN2024} line lists are also shown in the respective panels. A full list of the C$_2$H$_2$ line positions obtained in this work is provided in the supplementary table.

%%%%%%%%%%%%%%%%%%%%%%%%%%%%%%%%%%%%%%%%%
\section{Inverse Graph Construction of HDO energy levels}

\begin{figure}
\centering\includegraphics[width=0.8\textwidth]{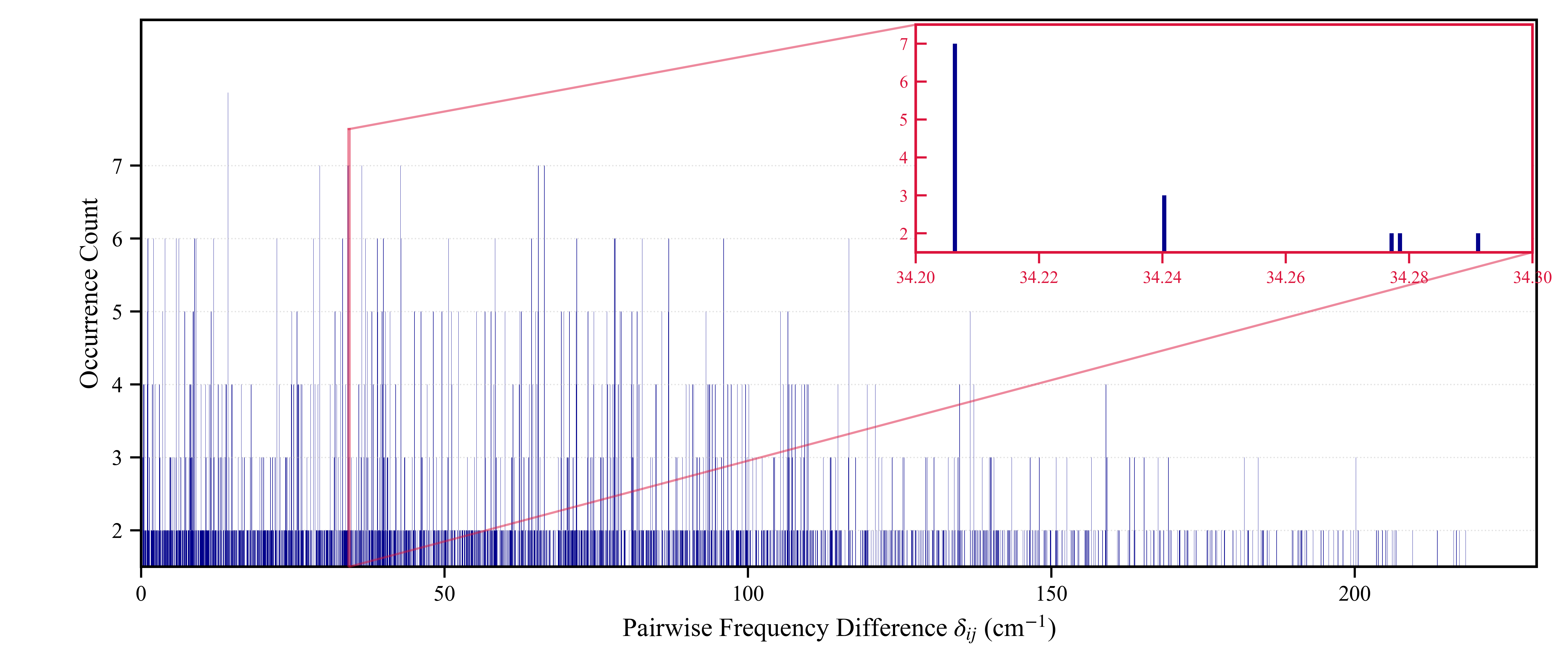}
\caption{Occurrence-count distribution of the pairwise frequency differences $\delta_{ij}=|f_i-f_j|$ obtained from the 686 water isotopologue Lamb dips. Recurring differences (occurrence count $N \ge 2$) are identified as candidate hyperedges for the inverse graph construction. Although over 2000 hyperedges with $N \ge 2$ were found, since each has a span of only $\sim 5\times 10^{-6}$~cm$^{-1}$, all hyperedges are well resolved, as shown in the inset. 
}
    \label{fig:count-delta}
\end{figure}

From the SCALS spectrum of HDO recorded in the range of $6230-6505$~cm$^{-1}$, we obtained a total of 686 Lamb dips, consisting of the set $\{f_i\}$.
The set $\{\delta_{ij} = |f_i-f_j|\}$ containing differences among all lines is obtained. Many of these values have a ``cluster'' feature. By taking a tolerance $\Delta$ value of $5\times 10^{-6}$~cm$^{-1}$ (roughly twice the experimental uncertainty), we count the occurrence of each $\delta_k$ value, and the result is plotted as Fig.~\ref{fig:count-delta}.
Note that a $\delta_k$ value with occurrence $N_k\ge 2$ may correspond to a K$_{2,2}$ motif shown in Fig.~\ref{fig:ARalgorithm}(c). Fig.~\ref{fig:count-delta} shows 2042 hyperedges (1679, 176, 104, 50, 25, 7, and 1 for $N=2, 3, \cdots 8$, respectively) found within a range of 270~cm$^{-1}$.
As each hyperedge has a width of only $\sim 5\times 10^{-6}$~cm$^{-1}$, all hyperedges are well resolved, and the chance of accidental occurrence has been effectively eliminated. 
We can see from the inset of Fig.~\ref{fig:count-delta} that this strategy only works when the resolution is sufficiently high. If the experimental accuracy drops to 0.001~cm$^{-1}$ (in most Doppler-limited studies), nearby hyperedges merge, and the algorithm can never work.

Therefore, a transition (edge) present in multiple hyperedges must be true in a K$_{2,2}$ motif. We find a transition (``edge'' in the graph) most shared by multiple motifs, and set it as an ``anchor'' for vertex propagation. Then we can reconstruct the whole network via propagation (see main text). 

\textbf{Level validation against MARVEL and HITRAN.} Of the 686 measured transitions, 425 are assigned to the reconstructed HDO network, and comparison with the HITRAN database~\cite{GORDON2026HITRAN2024} shows that the remaining 261 unassigned transitions arise predominantly from HD$^{16}$O (177) and D$_2^{16}$O (82), and two lines from H$_2^{16}$O. Nearly 80\% of the unassigned HD$^{16}$O lines fall in the lower-frequency region, pointing to insufficient coverage at higher frequencies due to limited laser tuning range; as a result, links between the K$_{2,2}$ motifs for $J=8$ and $J=9$ cannot be established, preventing determination of the lower-state energies of high-$J$ levels. The reconstructed levels agree with MARVEL~\cite{Tennyson2010} to within its uncertainties, except for eight levels whose deviations exceed 0.0005~cm$^{-1}$; the largest is 0.00256~cm$^{-1}$ for the $5_{4~1}$ [0~2~1] level at 6976.467500~cm$^{-1}$. In the ground state, the two largest deviations occur for the $4_{4,1}$ (402.328987~cm$^{-1}$) and $4_{4,0}$ (402.330812~cm$^{-1}$) levels, with deviations of $+0.00051$~cm$^{-1}$ and $-0.00059$~cm$^{-1}$, respectively. These two levels differ by only 0.002~cm$^{-1}$, so their transitions cannot be resolved in a Doppler-limited spectrum, which explains why their MARVEL energies, derived mainly from Doppler-broadened data, carry larger deviations. Six of the eight outliers correspond to states with $K_a = J$ or $J-1$, whose frequencies were previously derived from blended Doppler-limited spectra; their differences exceed the uncertainties claimed by MARVEL by factors of $6-27$.

%\spacing{1}
%\include{C2H2_assign_compared-c}
%\include{HDO-energy_data-drive}

\end{CJK*}

\end{document}